# Amorphous complexions alter the tensile failure of nanocrystalline Cu-Zr alloys

Jenna L. Wardini, Charlette M. Grigorian, Timothy J. Rupert [*]

Department of Materials Science and Engineering, University of California, Irvine, CA 92697, USA

* Email address: trupert@uci.edu

Grain boundary-based mechanisms are known to control the plastic deformation and failure of nanocrystalline metals, with manipulation of the boundary structure a promising path for tuning this response.  In this study, the role of interfacial structural disorder on plasticity and failure of nanocrystalline Cu-Zr alloys is investigated with in situ scanning electron microscopy tensile deformation experiments.  Two model materials are created, one with only the typical ordered grain boundaries and another with amorphous intergranular films interspersed into the boundary network, while the microstructures are otherwise identical.  Hence, the importance of complexion type on plasticity and failure is isolated by only varying complexion structure.  The tensile experiments show that failure of the samples containing amorphous films is significantly retarded, as evidenced by an increase in the cross-sectional area reduction, a decrease in the occurrence of shear-dominated failure, a decrease in strain localization, and fracture surfaces with more elongated dimple features.  As a whole, this study provides direct evidence that structural disorder at the grain boundaries can be beneficial for improving the ductility of nanocrystalline metals.

# 1. Introduction

Grain size refinement is a powerful approach for increasing the strength of metals and alloys for use in structural applications. As grain sizes are reduced to the nanoscale regime (< 100 nm), metallic systems undergo a shift in the dominant deformation mechanisms which control plasticity. In coarse-grained metals, plastic deformation primarily occurs by intragranular dislocation slip, but as grain size is reduced the boundaries themselves take on an increasingly active role in deformation. In the case of nanocrystalline metals, grain boundaries participate in plasticity by the emission and absorption of dislocations or by atomic shuffling to accommodate grain boundary sliding. With regards to engineering performance, the activation of these new mechanisms come with benefits such as greatly enhanced strength [1, 2], hardness [3], wear [4, 5], and fatigue resistance [6] of nanocrystalline metals over coarse-grained systems.

A primary barrier for the practical usage of these ultra-strong nanocrystalline metals and alloys is the typical loss of tensile ductility that often accompanies grain refinement [7]. Reports of tensile ductility in nanocrystalline metals and alloys are typically greatly reduced compared to coarse-grained metals, with low values of strain-to-failure below 2-3% often found in the literature for face centered cubic metals [3]. Some studies in the literature have suggested that these reduced ductility values can be exacerbated by extrinsic factors such as the existence of processing flaws or non-standardized test geometries [7, 8]. However, such extrinsic effects do not cause this reduced ductility directly, as nanocrystalline metals exhibit plastic flow and are not truly brittle materials [9], but instead only act to expose the major pre-existing weakness of nanocrystalline metals: *a heightened propensity for localized plastic deformation and premature failure*. The inability of nanoscale grains to accumulate intra-grain dislocation density results in a low strain-hardening capacity, but also means that dislocations must be absorbed at the opposite grain boundary once they have traversed the grain interior. Bitzek et al. [10] demonstrated that dislocation absorption can be a precursor to damage nucleation, with the local grain boundary stress and structure altered after the absorption event during molecular dynamics simulations. This prediction is consistent with the in situ transmission electron microscopy (TEM) deformation experiments of Kumar et al. [11] on nanocrystalline Ni, where internal cracks were found to nucleate at such grain boundary sites. Pan and Rupert [12] isolated the mechanism of repeated dislocation absorption at a grain boundary, indeed finding that cracks can nucleate when the local strain brought by the dislocations is not accommodated in an efficient manner. Some nanocrystalline metals, specifically those with grain sizes of only a few nanometers, exhibit shear localization reminiscent of the behavior of metallic glasses in the form of shear banding (see, e.g., [13-15]), while other reports of strain localization and crack nucleation/growth are tied to unaccommodated grain boundary sliding [16].

The common thread of all of the above discussion is the isolation of the grain boundary region as a site for damage nucleation, suggesting that boundary state is critical to damage tolerance. In fact, the local structure and chemistry of grain boundaries has been found to have a dramatic influence on dislocation-boundary interactions and the subsequent damage nucleation. For this reason, grain boundary engineering, where certain types of boundaries are preferentially introduced into the grain boundary network, has been a useful tool for increasing the toughness of advanced metallic materials (see, e.g., Bechtle et al. [17]). As an extension of the idea that boundary structure can impact damage creation, amorphous intergranular films (AIFs) were found to be better able to accommodate repeated dislocation absorption as compared to ordered grain boundaries (OGBs) [18], delaying crack nucleation and also slowing crack propagation. These authors showed that shear transformation zone activation within the structurally disordered boundary could cause the incoming strain to be accommodated over a larger volume within an AIF. Micropillar compression and micropillar bending experiments by Khalajhedayati et al. [19] provided direct validation of this concept, with AIF-containing Cu-Zr alloys exhibiting homogeneous deformation and increased bending ductility when compared to similar alloys with only OGBs. Subsequent experiments on nanocrystalline Al alloy films [20], nanostructured high-entropy nanocomposites [21], and superlattice alloys [22] have provided additional evidence that structural disorder at the grain boundary can in fact be beneficial to mechanical damage resistance, a concept introduced in Ref. [18]. AIFs and OGBs can be thought of as different thermodynamically-distinct grain boundary states or *complexions* [23, 24], in this case notably different in the level of structural order, giving a design variable that can potentially be used to tune the failure response of nanocrystalline metals.

Although AIFs have been identified as beneficial interfacial structures, their impact on failure and strain localization during tensile loading has not been isolated to date. In this study, we perform in situ scanning electron microscopy (SEM) tensile testing of two nanocrystalline Cu-Zr alloys, one with and one without AIFs, where all other microstructural descriptors (chemical composition, grain size, texture, second phase precipitation, etc.) are kept constant. We find that the addition of AIFs significantly alters the tensile failure mode of the nanocrystalline alloys. Comparisons of the mechanical behavior of the two materials reveal a slight increase in the yield and ultimate tensile strength, but more importantly a substantial increase in the ductility as measured by cross-sectional area-reduction. Investigation of the fracture surfaces demonstrates higher levels of local plastic flow in AIF-containing samples, and a resistance to localized failure through shearing modes based on measurements of fracture plane angle. Evaluation of the spatial distribution of strain along the tensile samples provides additional evidence that the sample containing AIFs deforms more homogeneously and is more damage tolerant. As a whole, this work strongly supports the conclusions that grain boundary structure can be used as a tool to control mechanical failure, with amorphous complexions serving to ductilize nanocrystalline metals.

## 2. Materials and Methods

Nanocrystalline Cu-Zr alloy powders were fabricated by high energy ball-milling in a SPEX SamplePrep 8000M mill using hardened steel milling media with a 10:1 ball-to-powder ratio under an inert atmosphere (99.99% pure Ar). High purity elemental Cu (Alfa Aesar, 99.99%, -170 + 400 mesh) and Zr (Micron Metals, 99.7%, -50 mesh) powders were milled in the presence of 1 wt.% stearic acid, which is used as a processing control agent to reduce cold-welding. Powders were milled for 10 h to refine the grain size into the nanocrystalline regime and induce mixing of the alloy component elements into a solid solution, face centered cubic phase. The resulting as-milled powders were split into two batches and encapsulated in quartz tubes under vacuum to prevent oxidation during the following 1 h anneal performed at 950 °C (~98% of the melting temperature of Cu-3 at.% Zr [25]). This annealing treatment encourages segregation of the Zr dopants to the grain boundaries and generates Zr-doped pre-melted regions with amorphous structure at some grain boundaries [19, 26]. After annealing, one set of powder samples was rapidly quenched by dropping directly from the furnace into a water bath, which serves to freeze in the amorphous boundary structures for further testing. This specimen contains AIFs interspersed within the grain boundary network, so we refer to this sample as the *AIF sample* for the remainder of the paper. In contrast, another set of powders was slowly cooled to room temperature after the annealing treatment, allowing the boundary region to crystallize and return to the typical grain boundary structure. Since this sample contains only ordered grain boundaries, we hereafter refer to this set of powders as the ***OGB sample***. By controlling the cooling procedure alone, two model materials which only vary in terms of grain boundary structure are synthesized, as the dopant segregation states and microstructural descriptors (average grain size, grain size distribution, impurity phase fraction, processing flaws, and impurities) are constant, which allows for a direct comparison of the mechanical behavior of the two materials based on their grain boundary structures alone. Additional details on the processing and characterization of these powders have been described previously in Refs. [19, 26, 27], while the application of these types of powder samples for the fabrication of bulk nanocrystalline metals is outlined in Ref. [28].

The structure and chemical composition of the Cu-Zr powders were characterized by X-ray diffraction (XRD), as well as TEM and scanning TEM (STEM) techniques. XRD was performed on a Rigaku SmartLab X-ray diffractometer, operated at 40 kV and 44 mA with Cu Kα radiation and a 1D D/teX Ultra 250 detector, to identify secondary phases, secondary phase volume fractions, and average sizes (Figure 1(a) and Table 1). The average grain size of the Cu-rich face centered cubic matrix phase was ~70 nm, while impurity phases with mean grain sizes of ~60-90 nm were also identified. These second phases were found in relatively small amounts, with <0.6 vol.% for $ZrO_2$ and <2.3 vol.% for ZrC. Importantly, the size and fraction of these second phases did not vary significantly between the two sample sets. TEM lamellae were prepared using an FEI Quanta 3D dual-beam focused ion beam (FIB)/SEM, following established TEM sample preparation procedures [29]. TEM inspection was performed on a JEOL-2800 operated at 200 kV. Bright field (BF)-STEM was used to confirm the nanocrystalline grain sizes of the Cu matrix phases (Figure 1(d)), while high-angle annular dark-field (HAADF)-STEM images were collected simultaneously to ensure that Zr-rich secondary phases were not mistaken for Cu grains during grain size determination (Figures 1(b) and 1(c)).  Energy-dispersive X-ray spectroscopy (EDS) was performed on an FEI-Magellan 400 SEM to measure the average Zr concentration (3.43 ± 0.08 at.% Zr) and to investigate fracture surfaces using a 5 kV electron beam and in-lens detector with the immersion lens activated.

Micro-tensile samples were prepared on an FEI Quanta 3D Dual-Beam FIB/SEM using a semi-automated FIB lathe milling approach first developed by Uchic and Dimiduk [30].  Although this approach is somewhat time-intensive, it allows for a superior level of control over sample geometry and the creation of taper-free specimens to ensure that a homogeneous deformation is applied [31].  It is also the only method that delivers a cylindrical gauge section in situations where thin-film patterning approaches cannot be used, such as the testing on individual powder particles performed here.  The Cu-Zr powders were first embedded in a graphite-filled epoxy on an SEM stub, and then mechanically polished to expose both the top and side surface of powder particles (Figure 2(a)).  A 3-5 µm thick Pt cap is ion beam deposited onto the surface to protect the underlying material during the milling steps to follow.  A rough micropillar shape is first formed by removing an annulus of material to a depth of ~30 µm, first at a high milling current of 50-65 nA (Figure 2(b)) and then at a lower milling current of 7-15 nA as the inner annulus diameter approaches the Pt cap (Figure 2(c)).  The sample is then viewed from the side and the Pt cap is removed with the FIB to create a flat surface (Figure 2(d), black box).  A circular fiducial mark, chosen for its rotational symmetry, is then milled at 0.1 nA onto the top surface to facilitate automated pattern matching, as the sample is incrementally rotated and repositioned throughout the lathe milling process.

To begin the lathe milling process, the SEM stage is tilted to -8° with respect to the sample surface so the ion beam can impact the long axis of the pillar at 60° (as near to perpendicular as possible under instrument constraints). The stage is then rotated and pillar side is milled every 10-20° degrees, re-centering the pillar by pattern matching and centering of the fiducial marker before initiating milling at each step. A single pass of lathe milling implies the sample has been rotated through a full 360° rotation and the micro-tensile specimen preparation can usually be achieved in four or five passes of lathe-milling, depending on the desired level of shape refinement and surface polish. In the first pass, a rectangular milling pattern is used to remove any taper along the pillar side (Figure 2(d), red box), which allows for more predictable milling behavior in the sample-shaping passes to follow. In the second pass, the dogbone shape is created with a trapezoidal mill pattern at a current of 3-5 nA, with ion beam images at multiple rotation angles shown in Figure 2(e). The next to last pass refines the dogbone shape, correcting for any asymmetry and surface roughness left from the previous steps using a reduced milling current (0.5-1 nA) and a smaller milling area for more controlled shaping (Figure 2(f)). The final pass polishes the surface of the gauge section to remove any surface roughness, and is usually performed at a current of 0.1 nA or below. An example of a finished micro-tensile specimen can be seen in Figure 2(g).

A FemtoTools nanomechancial testing system (model FT-NMT03, FemtoTools, Buchs, Switzerland) was used to conduct micro-tension tests under SEM observation. A $50 \times 50$ $\mu m^2$ flat Si MEMS-based microforce sensor head (model FT-S200'000) was milled into a micro-tensile grip using the FIB (Figure 3(a)). The angles and dimensions of the grip were chosen to match the micro-tensile samples as closely as possible to provide maximum contact area between the grip and the test piece. During the experiments, the displacement of the tensile grip was controlled with a sub-nanometer resolution piezo-based actuation system. All micro-tensile tests were conducted in displacement-controlled mode at a nominal strain rate of $10^{-3}$ $s^{-1}$ and at room-temperature. Actual strain rates varied slightly during tension testing due to the nonlinear compliance of the testing system under tension in the initial stages of loading. To improve the resolution and reliability of strain calculations, measurements directly on the gauge section were made through frame-by-frame tracking of electron-beam deposited Pt gauge markers (Figures 3(b) and 3(c)). The temporal resolution of these strain measurements was limited to 2 Hz, based on the fastest SEM frame acquisition rate available in the FEI software for video recording. The spatial resolution of marker position measurements was controlled by a combination of pixel size of the acquired frame, Pt marker size (estimated to be between 30-80 nm across experiments), and fidelity of the Pt marker atop the dynamically deforming background of the tensile surface. The engineering strain across the gauge section was extracted using a custom MATLAB© script designed to locate the position of Pt markers based on the maximum pixel intensity integrated along the marker's length, perpendicular to the tensile axis. Error in Pt marker position in the latter part of tests is primarily due to breakdown of the Pt marker due to extensive deformation, which creates offsets in its average position. Image drift was assessed to be below the spatial detection limit for all tests conducted. A simplified version of the tensile sample fabrication and mechanical testing can be found in Supplementary Video 1.

Two types of tensile samples were fabricated: (1) a smaller diameter ($D$) set (eight samples, $D$ = 1.9 - 4 µm) with two gauge markers to track global strain (Figure 3(b)), and (2) a larger diameter set (four samples, $D$ = 4.7 - 5.6 µm) with nine gauge markers along the length to track both global and local strains (Figure 3(c)). It is important to note that all pillar diameters were much larger than the grain size of the nanocrystalline alloy being studied in an attempt to avoid external size effects on the measured mechanical properties [32, 33]. The larger diameter sample set was fabricated with the goal of reducing the potential impact of surface flaws on the reproducibility of strain-to-failure values [8], as well as minimizing any potential impact of $Ga^+$ ion surface penetration on the mechanical behavior by increasing the volume-to-surface area ratio. In addition, the larger sample geometry made it possible to deposit more markers along the gauge length to enable local strain measurements. The micro-tensile sample aspect ratios were between 3 and 5 for all specimens tested in this study.

## 3. Results and Discussion

Stress-strain curves collected by the in situ micro-tensile testing experiments were first collected for five OGB samples and seven AIF samples. Overall sample strain, or global strain, is measured from the two outermost markers on the gauge section, providing an averaged strain measurement from the entire sample gauge section. Representative examples of the stress-strain behavior for each sample type are presented in Figure 4. The morphological evolution of the gauge sections as the samples begin to yield and then progress to failure are shown in a series of SEM images, with the corresponding points on the stress-strain curve labeled. Figures 4(b)-4(f) show that after the OGB sample achieves its ultimate tensile strength, further plastic deformation is no longer homogeneous and becomes localized along a shear plane (Supplementary Video 2). Shear off-sets can be seen on opposite sides of the sample surface, first becoming apparent in Figure 4(d). At this point, the sample continues to hold load, and further displacement of the piezo-stage causes the top and bottom sections of the test piece to slide with respect to one another along the shear plane until they disconnect. Interestingly, the shear failure plane appears to connect two small surface pores on opposite sides of the gauge diameter, demonstrating that OGB samples are highly sensitive to such defects. In contrast, the plastic flow after the ultimate strength in the AIF sample (Figures 4(j)-4(l)) manifests as a diffuse and symmetrical reduction in the cross-section through classical necking behavior (Supplementary Video 3). Although some of the samples exhibit a mixed mode failure behavior where it was difficult to ascribe one failure mode or the other, generally the shear-type abrupt failure dominated in OGB samples while the diffuse and symmetric necking behavior dominated in AIF samples.

The compiled stress-strain curves of all OGB and AIF samples are presented in Figure 5. Both sample types demonstrate relatively high strength in comparison with pure nanocrystalline Cu with a similar grain size [34]. This enhanced strength is primarily due to Zr dopant segregation to the grain boundaries, following prior reports of stronger nanocrystalline alloys compared to pure metals with equivalent grain structures [35]. The stress-strain curves also display other features common to nanocrystalline metals such as minimal strain-hardening and a softening behavior after reaching the ultimate strength. This softening can be attributed to a number of causes, such as a reduction in cross-sectional area at the neck (Figures 4(h)-4(l) for an AIF sample), the formation of a plane of localized shear (Figures 4(b)-4(f) for an OGB sample), and/or the reduction of load bearing area due to macroscale crack formation and opening (to be shown later in Figure 11 for OGB samples). Stress-strain curves were truncated at the point where there is either a vertical drop in stress of 100 MPa or greater, or when the sample has visibly separated due to crack opening.

Figure 5(c) summarizes the results from the twelve tensile experiments, with the ultimate tensile strength and strain-to-failure reported for each sample. In this figure, the shaded boxes indicate one standard deviation from the average ultimate tensile strength and strain-to-failure values, which are indicated by horizontal or vertical lines, respectively. These averages leave out results from two tensile samples, one from each sample set, with the smallest diameters ($D = 1.86$ μm for the OGB sample and $D = 2.12$ μm for the AIF sample), as indicated by open data points in Figure 5(c). It has been well-established that micropillars can exhibit mechanical properties that depend on the external size of the sample (see, e.g., Ref. [36]). However, this behavior is typically observed in single crystal samples. A micron-sized sample can still be large enough to contain a representative volume element of material provided the scale associated with microstructural features is much smaller than the pillar dimensions. For nanocrystalline materials, this means that the average grain size must be much smaller than the pillar diameter. Gu et al. [33] and Jang and Greer [32] explored this phenomenon as it relates to specimen strength in nanocrystalline Pt and Ni-W, respectively, finding that external size effects can be avoided if the pillar diameter are at least 25-30 times larger than the average grain size. Since the two thinnest samples were not clearly above this threshold, we have reason to believe that an external size effect could be active. Gu et al. [33] demonstrated a reduction in strength with decreasing pillar diameter that was the result of having a large number of grains exposed to a free surface when the pillar is very small. In that case, deformation is proposed to be controlled by grain boundary sliding rather than by dislocation activity due to the relaxed boundary conditions at the pillar surface. Consistent with these prior reports, our two smallest samples had relatively low strengths compared to the rest of the specimens for a given sample group, providing another signal that an external size effect may be active and this data should be excluded. We also find that narrow samples, especially those with a greater aspect ratio, are more affected by test misalignment. For example, the smallest diameter AIF sample was observed to undergo a slight lateral bending deformation due to misalignment during the experiment. For all of these reasons, we exclude these two thinnest samples from our subsequent analysis, but include them in this figure in the name of transparency and completeness. We note that the specimens shown previously in Figure 4 both had acceptably large pillar diameters to avoid any external size effects.

The average ultimate tensile strength values were calculated to be 767 ± 47 MPa for the OGB samples and 805 ± 52 MPa for the AIF samples, indicating a mild strength enhancement of strength (~5%) for the AIF sample. We reiterate that the grain size, segregation state, and carbide size and distribution is the same between the OGB and AIF sample sets, so this strengthening effect can be attributed to the grain boundary structure. Khalajhedayati et al. [19] also measured a strengthening increment in AIF samples with microcompression testing, so our results provide evidence that the strengthening effect is also found in tension. Turlo and Rupert [37] used atomistic simulations to investigate possible mechanistic explanations for such a strengthening effect from AIFs as compared to OGBs, isolating the pinning of dislocations along the grain sides during propagation across the crystal as the critical event. These authors found that AIFs increase nanocrystalline strength due to a combination of local stress variations and local ledges in the boundary, both of which act as stronger pinning sites.

The average strain-to-failure values were nearly the same for the two samples, with 8.0 ± 2.5% measured in the OGB sample and 7.7 ± 3.2 % in the AIF sample. We note that the specimen with the highest strain-to-failure of 13.2% was an AIF sample, compared with the best performer from the OGB samples that had a strain-to-failure of 10.3%. To understand the plasticity of each sample set more clearly, the fracture surfaces were inspected in detail. Figures 6(a) and (b) show representative examples of fracture surfaces from OGB and AIF samples. Both surfaces display dimpling, commonly associated with plastic flow and the intrinsic or metallurgical ductility expected of even a nanocrystalline metal. However, in the OGB sample, the material drawn around the dimples has a lower aspect ratio, signaling more limited local flow. Generally, the OGB samples displayed a more planar failure surface covered in parabolic dimpling that follows the direction of failure. In contrast, a prevalent feature observed for the AIF fracture surfaces was a complex and turbulent topography which did not easily fit to a plane, suggesting that a multiaxial stress field associated with plastic rupture and void coalescence was active in this set of samples.

The AIF samples appear to be intrinsically more ductile based on inspection of the fracture surfaces, yet the strain-to-failure measurements do not provide conclusive evidence to support one set of samples or another. To understand this apparent discrepancy, it is necessary to consider if one observation or the other could be flawed for some reason. As noted by Brooks et al. [8] in their study of the tensile behavior of millimeter-scale electrodeposited nanocrystalline Ni films, processing defects such as porosity can result in low reproducibility and reliability of strain-to-failure values. For the materials studied here, processing defects such as pores created during the ball milling process or impurity phases such as oxides and carbides could affect the direct measurement of strain-to-failure values. Figure 7(a) shows the fracture surface of a large diameter sample ($D$ = 5.6 µm) while Figure 7(b) shows the facture surface of a small diameter sample ($D$ = 2.8 µm) presented at the same magnification, with both images coming from the AIF-containing sample set. Obvious processing defects are outlined with white dashed lines. In Figure 7(a), the individual processing defects were relatively small compared to the sample size, and there is the highly dimpled band of material running diagonal through the cross-section which contains a much lower density of defects than the surrounding material. Here, it appears that the defect-free region carried the load after crack initiation, as evidence of extensive plasticity is found in this region. In contrast, the sample shown in Figure 7(b) failed at a low applied strain without signs of extensive local deformation on the fracture surface, most likely because it happened to have an extremely large pore in the gauge section, shown at the top left of the fracture surface. Figure 7(c) presents a magnified image of a fracture surface, showing a multi-scale processing defect structure where both nanoscale impurity particles and microscale pores can be observed. Strain-to-failure measurements are extremely sensitive to these types of processing defects, suggesting that such measurements are not reliable indicators of ductility for these samples.

While strain-to-failure is not a reliable representation of ductility in this study due to processing defects, area reduction is another well-established metric that is regularly used to quantify ductility. The reduction in cross-sectional area for all specimens was measured by SEM inspection of their fracture surfaces, viewed normal to the tensile axis. Two representative examples of fracture surfaces from each sample type are shown in Figures 8(a)-(d). In these images, the OGB samples show area reductions of 12% (Figure 8(a)) and 6% (Figure 8(b)), while the AIF samples demonstrate more obvious neck formation with area reductions of 45% (Figure 8(c)) and 26% (Figure 8(d)). Figure 8(e) presents a summary of the area-reduction across the entire data set, where mean values are indicated with solid horizontal lines and standard deviations with shaded rectangles. The OGB samples have an average area reduction of 18%, while the AIF samples have an average area reduction of 32%. Based on these measurements, we can conclude that there is a 44% relative increase in ductility when AIFs are incorporated into the nanocrystalline grain boundary network. This difference in ductility can be further shown by inspecting the gauge sections from the side, or perpendicular to the tensile axis. The OGB sample presented in Figure 8(f), which is the same specimen shown in Figure 8(b), has little to no shape change in the gauge section, with the strain highly localized along a plane tilted with respect to the tensile axis. In contrast, the AIF sample presented in Figure 8(g), which is the same specimen shown in Figure 8(c), has an obvious necked region with a significantly reduced diameter.

A number of specimens experienced shear banding, or the localization of plastic strain into a distinct shear plane. To quantify the tendency to shear band, a planar surface was fitted to the fracture surface from a side-view so that the angle with respect to the tensile loading axis could be measured. Figure 9(a) shows an extreme case of shear band failure for an OGB sample, with a measured fracture angle of ~45°. Figure 9(b) shows a representative example from an AIF sample, where the fracture plane is closer to horizontal. We note in this figure that there is a slight incline of the fracture plane with respect to the viewing direction, which is somewhat difficult to see directly in Figure 9(b). The angle of the fracture plane was measured to be ~23° after the sample was removed from the nanomechanical testing system and remounted. For all samples, the inclination angle of the fracture plane of both the base and detached top piece remaining in the grip were measured and the reported values are the average of these two values. The fracture angle measurements are summarized in Figure 9(c), indicating that on average the OGB fracture surfaces have a higher inclination angle with respect to the tensile axis and are more likely to localize deformation along a shear band. Shear banding is a common deformation mode of metallic glasses [38, 39], and Trelewicz and Schuh [13, 14] showed that the finest nanocrystalline grain sizes can experience shear banding as well. Both simulations [40] and experiments [15] have shown that this type of localized deformation can lead to grain growth within the shear band. Khalajhedayati and Rupert [41] used microcompression to show that grain boundary state can affect nanocrystalline shear banding, with relaxed and structurally ordered boundaries more likely to experience this failure mode. Our observation that shear banding is at least partially suppressed in the samples with AIFs provides additional support for the concept that structural disorder at the grain boundaries can be beneficial for mechanical properties.

Finally, since our tests are performed in situ, we can investigate the local strain along the sample gauge section during the experiments. Local strain ($\varepsilon_{local}$) distributions along the tensile axis were tracked for the four samples (two OGB and two AIF) with nine Pt fiducial markers deposited along the gauge section, separating the gauge length into eight equally spaced segments.

Figure 10 presents the results of this analysis, where (a)-(d) show in situ SEM images of the gauge sections for each of the four samples tested at global strains ($\varepsilon_{global}$) of 0.0% (undeformed) and 7.2%. The strain value of 7.2% was chosen because it is the highest global strain achieved by the lowest performer of the four, representing the highest plastic strain that is still comparable across all samples. $\varepsilon_{local}$ refers to the local strain across one segment, while $\varepsilon_{local\ max}$ is the maximum local strain achieved by each of the eight segments along the tensile axis at 7.2% global strain. The ratio of $\varepsilon_{local}/\varepsilon_{local\ max}$ then provides a normalized metric by which the relative strain distributions can be compared. Figures 10(e)-(h) display this ratio projected onto the deformed sample images via a colored strain map. Figures 10(i)-(l) then plot this same information as strain distribution profiles, where Segment 1 is the left most segment representing the base of the tensile test piece and Segment 8 indicates the right most segment, or the top of the tensile test piece. Both OGB samples show that strain is highly localized to a single segment along the gauge section. The maximum local strain is at least two times larger than the local strain anywhere else along the gauge length, while the maximum local strain is five times higher than the local strain elsewhere for the second OGB sample. In contrast, the AIF specimens exhibit a more evenly distributed strain profile, with highly strained regions found distributed along the gauge length.

The local strain analysis for these samples is extended to the instant immediately prior to failure in Figure 11, with the local strain values now shown in absolute terms. It is apparent that both of the OGB samples form visible cracks at either internal or surface flaws before complete separation, which results in highly localized strain near the eventual failure point. In contrast, the AIF samples show no signs of crack formation, but fail after the formation of a diffuse neck. The two surface flaws on the specimen shown in Figure 11(d) where present in the undeformed configuration shown in Figure 10(d), proving a level of flaw tolerance for the AIF-containing samples that is not evident in the OGB samples. The reduction in strain localization in the AIF sample is again clearly shown by the spatial distribution of the local strain, which is broad and evenly distributed in Figures 11(g) and 11(h). As a whole, both Figures 10 and 11 demonstrate that the AIF samples have the ability to plastically deform in a more homogeneous manner than the OGB samples.

The localization of plastic strain in pure nanocrystalline metals is proposed to occur in three main stages [40]. First, local shear strains concentrate at grain boundaries under tensile loading prior to the onset of inhomogeneous plastic flow. These stress concentrations can occur due to either inefficient dislocation absorption [12] or when grain boundary sliding which is frustrated [33, 42]. Next, a shear path composed of interconnected and highly strained grain boundaries can form across the sample diameter and becomes a preferential site for further deformation. Finally, plastic localization to the shear path can be intensified by stress-driven grain coarsening, which further weakens this region relative to the bulk. At this stage, the limited strain hardening capability of nanocrystalline metals leaves them largely defenseless to runaway plasticity. With continued loading, the shear path either widens into a shear band and/or locally weakens the specimen until it is overloaded.

AIFs are effective in mitigating strain localization by actively impeding the formation of localized shear bands at each of these stages. AIFs suppress the development of the long-range shear strain paths that precede shear band formation, and they inhibit mechanisms that accelerate inhomogeneous plastic flow once localization has initiated. The disordered structure and increased thickness of AIFs allows local plastic strain to be accommodated over a larger volume (i.e., less concentrated) within the grain boundary region during dislocation absorption [18]. Measurements of the axial strain distribution during the micro-tensile tests conducted here directly demonstrate the ability of AIFs to effectively resist localization of strain on the macroscale. Once a stress concentration is created and a crack eventually forms, the AIFs can slow propagation of the defect through the sample, as seen in Figure 11. Atomistic modeling results from Pal et al. [43] suggested that AIFs can resist propagation of both parallel and perpendicular cracks by blunting the crack tip. In situ TEM fatigue experiments by Schuler et al. [44] provided confirmation of such crack tip blunting by AIF-containing microstructures, as well as distribution of plastic activity into a larger region inside the material. AIFs will also dampen localization progression by inhibiting grain coarsening that leads to localized softening in the shear path, as amorphous grain boundary complexions have been observed to have a stabilizing effect on grain size [26, 45]. The enhanced grain size stability observed in AIF containing nanocrystalline binary alloys under thermal treatments should be mirrored under mechanical drivers for grain coarsening, as the fundamental principle of reducing the driving force for grain boundary migration by reducing the grain boundary free energy is the same.

AIFs represent a tunable microstructural feature as well, as higher temperature heat treatments give thicker amorphous films [46, 47], with recent experiments showing that quench rate can also be manipulated to tune the thickness of amorphous complexions in a nanocrystalline grain structure [48]. Atomistic modeling studies have shown that an AIF's ability to blunt cracks [43], slow crack propagation [18, 43], and absorb dislocations [18] increases with increasing thickness. Therefore, it is not only beneficial to have AIFs within a grain boundary network, but it is also possible to vary the thickness of these complexions to increase damage tolerance.

## 4. Summary and Conclusions

In this study, the tensile deformation and failure was investigated for two Cu-Zr alloys whose only difference is the degree of structural ordering at the grain boundaries. Differences in deformation behavior were assessed through their tensile stress-strain behavior, as well as features of the deformed tensile gauge sections and fracture surfaces. Various metrics are used to demonstrate that nanocrystalline ductility can be increased by AIFs including area reduction, fracture angle, and the distribution of strain along the tensile axis during the deformation experiments. From these analyses, we conclude that:

- While strain-to-failure measurements are inconclusive due to the effect of processing defects, a 44% increase in area reduction for the AIF samples demonstrates improved ductility as compared to the OGB samples. A greater ability to sustain a diffuse and well-formed neck under tensile load was observed for the specimens containing AIFs.

- AIF samples show an increased resistance to localized, shear-type failure. In contrast, OGB samples were more likely to experience catastrophic failure due to shear banding.

- The local strain distributions, measured in situ during the tensile experiments, show that AIF samples experience more homogeneous plastic deformation along the gauge section, eventually failing due to void formation with fracture surfaces containing evidence of high levels of local plasticity. In contrast, OGB samples heavily localize plastic strain. It is hypothesized that the more homogeneous macroscopic strain distribution arises from the ability of AIFs to prevent extreme strain localization at the atomistic scale, at and in the vicinity of grain boundaries.

As a whole, this study provides direct experimental evidence that common failure processes such as shear banding and crack formation arising from extreme strain localization in nanocrystalline metals can be mitigated by altering grain boundary structure. Varying this structure, namely increasing the structural disorder of the boundaries, can change the way that incoming dislocations interact with grain boundaries and improve the ductility of nanocrystalline alloys.

## Acknowledgments

This study was supported by the U.S. Department of Energy, Office of Basic Energy Sciences, Materials Science and Engineering Division under Award No. DE-SC0021224.

# References

[1] K.M. Youssef, R.O. Scattergood, K.L. Murty, J.A. Horton, C.C. Koch, Ultrahigh strength and high ductility of bulk nanocrystalline copper, Appl. Phys. Lett. 87(9) (2005) 091904.
[2] J.R. Weertman, Hall-Petch Strengthening in Nanocrystalline Metals, Mater. Sci. Eng. A 166(1-2) (1993) 161-167.
[3] C.C. Koch, K.M. Youssef, R.O. Scattergood, K.L. Murty, Breakthroughs in Optimization of Mechanical Properties of Nanostructured Metals and Alloys, Adv. Eng. Mater. 7(9) (2005) 787-794.
[4] T.J. Rupert, W. Cai, C.A. Schuh, Abrasive wear response of nanocrystalline Ni–W alloys across the Hall–Petch breakdown, Wear 298–299(0) (2013) 120-126.
[5] T.J. Rupert, C.A. Schuh, Sliding wear of nanocrystalline Ni-W: Structural evolution and the apparent breakdown of Archard scaling, Acta Mater. 58(12) (2010) 4137-4148.
[6] N.M. Heckman, S.M. Foiles, C.J. O'Brien, M. Chandross, C.M. Barr, N. Argibay, K. Hattar, P. Lu, D.P. Adams, B.L. Boyce, New nanoscale toughening mechanisms mitigate embrittlement in binary nanocrystalline alloys, Nanoscale 10(45) (2018) 21231-21243.
[7] M. Dao, L. Lu, R.J. Asaro, J.T.M. De Hosson, E. Ma, Toward a quantitative understanding of mechanical behavior of nanocrystalline metals, Acta Mater. 55(12) (2007) 4041-4065.
[8] I. Brooks, G. Palumbo, G.D. Hibbard, Z.R. Wang, U. Erb, On the intrinsic ductility of electrodeposited nanocrystalline metals, J. Mater. Sci. 46(24) (2011) 7713-7724.
[9] J.A. Sharon, H.A.I. Padilla, B.L. Boyce, Interpreting the ductility of nanocrystalline metals, J. Mater. Res. 28(12) (2013) 1539-1552.
[10] E. Bitzek, C. Brandl, D. Weygand, P.M. Derlet, H. Van Swygenhoven, Atomistic simulation of a dislocation shear loop interacting with grain boundaries in nanocrystalline aluminium, Modell. Simul. Mater. Sci. Eng. 17(5) (2009) 055008.
[11] K.S. Kumar, S. Suresh, M.F. Chisholm, J.A. Horton, P. Wang, Deformation of electrodeposited nanocrystalline nickel, Acta Mater. 51(2) (2003) 387-405.
[12] Z. Pan, T.J. Rupert, Damage nucleation from repeated dislocation absorption at a grain boundary, Comput. Mater. Sci. 93(0) (2014) 206-209.
[13] J.R. Trelewicz, C.A. Schuh, The Hall-Petch breakdown in nanocrystalline metals: A crossover to glass-like deformation, Acta Mater. 55(17) (2007) 5948-5958.
[14] J.R. Trelewicz, C.A. Schuh, The Hall-Petch breakdown at high strain rates: Optimizing nanocrystalline grain size for impact applications, Appl. Phys. Lett. 93(17) (2008) 171916.
[15] A. Khalajhedayati, T.J. Rupert, Disruption of Thermally-Stable Nanoscale Grain Structures by Strain Localization, Scientific Reports 5 (2015) 10663.
[16] E. Hosseinian, S. Gupta, O.N. Pierron, M. Legros, Size effects on intergranular crack growth mechanisms in ultrathin nanocrystalline gold free-standing films, Acta Mater. 143 (2018) 77-87.
[17] S. Bechtle, M. Kumar, B.P. Somerday, M.E. Launey, R.O. Ritchie, Grain-boundary engineering markedly reduces susceptibility to intergranular hydrogen embrittlement in metallic materials, Acta Mater. 57(14) (2009) 4148-4157.
[18] Z. Pan, T.J. Rupert, Amorphous intergranular films as toughening structural features, Acta Mater. 89(0) (2015) 205-214.
[19] A. Khalajhedayati, Z. Pan, T.J. Rupert, Manipulating the interfacial structure of nanomaterials to achieve a unique combination of strength and ductility, Nature Communications 7 (2016) 10802.

[20] G. Wu, C. Liu, L. Sun, Q. Wang, B. Sun, B. Han, J.-J. Kai, J. Luan, C.T. Liu, K. Cao, Y. Lu, L. Cheng, J. Lu, Hierarchical nanostructured aluminum alloy with ultrahigh strength and large plasticity, Nature Communications 10(1) (2019) 5099.
[21] G. Wu, S. Balachandran, B. Gault, W. Xia, C. Liu, Z. Rao, Y. Wei, S. Liu, J. Lu, M. Herbig, W. Lu, G. Dehm, Z. Li, D. Raabe, Crystal–Glass High-Entropy Nanocomposites with Near Theoretical Compressive Strength and Large Deformability, Adv. Mater. 32(34) (2020) 2002619.
[22] T. Yang, Y.L. Zhao, W.P. Li, C.Y. Yu, J.H. Luan, D.Y. Lin, L. Fan, Z.B. Jiao, W.H. Liu, X.J. Liu, J.J. Kai, J.C. Huang, C.T. Liu, Ultrahigh-strength and ductile superlattice alloys with nanoscale disordered interfaces, Science 369(6502) (2020) 427-432.
[23] P.R. Cantwell, T. Frolov, T.J. Rupert, A.R. Krause, C.J. Marvel, G.S. Rohrer, J.M. Rickman, M.P. Harmer, Grain Boundary Complexion Transitions, Annual Review of Materials Research 50(1) (2020) 465-492.
[24] P.R. Cantwell, M. Tang, S.J. Dillon, J. Luo, G.S. Rohrer, M.P. Harmer, Grain boundary complexions, Acta Mater. 62(0) (2014) 1-48.
[25] H. Okamoto, Cu-Zr (Copper-Zirconium), Journal of Phase Equilibria and Diffusion 29(2) (2008) 204-204.
[26] A. Khalajhedayati, T.J. Rupert, High-Temperature Stability and Grain Boundary Complexion Formation in a Nanocrystalline Cu-Zr Alloy, JOM 67(12) (2015) 2788-2801.
[27] C.M. Grigorian, T.J. Rupert, Thick amorphous complexion formation and extreme thermal stability in ternary nanocrystalline Cu-Zr-Hf alloys, Acta Mater. 179 (2019) 172-182.
[28] O.K. Donaldson, T.J. Rupert, Amorphous Intergranular Films Enable the Creation of Bulk Nanocrystalline Cu–Zr with Full Density, Adv. Eng. Mater. 21(9) (2019) 1900333.
[29] M.H.F. Overwijk, F.C.v.d. Heuvel, C.W.T. Bulle-Lieuwma, Novel scheme for the preparation of transmission electron microscopy specimens with a focused ion beam, Journal of Vacuum Science & Technology B 11(6) (1993) 2021-2024.
[30] M.D. Uchic, D.M. Dimiduk, A methodology to investigate size scale effects in crystalline plasticity using uniaxial compression testing, Materials Science and Engineering: A 400-401 (2005) 268-278.
[31] H. Zhang, B.E. Schuster, Q. Wei, K.T. Ramesh, The design of accurate micro-compression experiments, Scr. Mater. 54(2) (2006) 181-186.
[32] D.C. Jang, J.R. Greer, Size-induced weakening and grain boundary-assisted deformation in 60 nm grained Ni nanopillars, Scr. Mater. 64(1) (2011) 77-80.
[33] X.W. Gu, C.N. Loynachan, Z.X. Wu, Y.W. Zhang, D.J. Srolovitz, J.R. Greer, Size-Dependent Deformation of Nanocrystalline Pt Nanopillars, Nano Lett. 12(12) (2012) 6385-6392.
[34] M.A. Meyers, A. Mishra, D.J. Benson, Mechanical properties of nanocrystalline materials, Prog. Mater. Sci. 51(4) (2006) 427-556.
[35] S. Ozerinc, K. Tai, N.Q. Vo, P. Bellon, R.S. Averback, W.P. King, Grain boundary doping strengthens nanocrystalline copper alloys, Scr. Mater. 67(7-8) (2012) 720-723.
[36] J.R. Greer, J.T.M. De Hosson, Plasticity in small-sized metallic systems: Intrinsic versus extrinsic size effect, Prog. Mater. Sci. 56(6) (2011) 654-724.
[37] V. Turlo, T.J. Rupert, Grain boundary complexions and the strength of nanocrystalline metals: Dislocation emission and propagation, Acta Mater. 151 (2018) 100-111.
[38] C.A. Schuh, T.C. Hufnagel, U. Ramamurty, Mechanical behavior of amorphous alloys, Acta Mater. 55(12) (2007) 4067-4109.

[39] A.L. Greer, Y.Q. Cheng, E. Ma, Shear bands in metallic glasses, Materials Science and Engineering: R: Reports 74(4) (2013) 71-132.
[40] T.J. Rupert, Strain localization in a nanocrystalline metal: Atomic mechanisms and the effect of testing conditions, J. Appl. Phys. 114(3) (2013).
[41] A. Khalajhedayati, T.J. Rupert, Emergence of localized plasticity and failure through shear banding during microcompression of a nanocrystalline alloy, Acta Mater. 65 (2014) 326-337.
[42] X.W. Gu, Z. Wu, Y.W. Zhang, D.J. Srolovitz, J.R. Greer, Microstructure versus Flaw: Mechanisms of Failure and Strength in Nanostructures, Nano Lett. 13(11) (2013) 5703-5709.
[43] S. Pal, K. Vijay Reddy, C. Deng, On the role of Cu-Zr amorphous intergranular films on crack growth retardation in nanocrystalline Cu during monotonic and cyclic loading conditions, Comput. Mater. Sci. 169 (2019) 109122.
[44] J.D. Schuler, C.M. Barr, N.M. Heckman, G. Copeland, B.L. Boyce, K. Hattar, T.J. Rupert, In Situ High-Cycle Fatigue Reveals Importance of Grain Boundary Structure in Nanocrystalline Cu-Zr, JOM 71(4) (2019) 1221-1232.
[45] J.D. Schuler, O.K. Donaldson, T.J. Rupert, Amorphous complexions enable a new region of high temperature stability in nanocrystalline Ni-W, Scr. Mater. 154 (2018) 49-53.
[46] J. Luo, V.K. Gupta, D.H. Yoon, H.M. Meyer, Segregation-induced grain boundary premelting in nickel-doped tungsten, Appl. Phys. Lett. 87(23) (2005).
[47] J. Luo, X.M. Shi, Grain boundary disordering in binary alloys, Appl. Phys. Lett. 92(10) (2008).
[48] C.M. Grigorian, T.J. Rupert, Critical cooling rates for amorphous-to-ordered complexion transitions in Cu-rich nanocrystalline alloys, Acta Mater. 206 (2021) 116650.

**Table 1.** XRD volume fractions and grain sizes for the primary face centered cubic Cu-rich phase, as well as impurity oxide and carbide phases.

|     | Vol. % | | Average Grain Size | | |
| --- | --- | --- | --- | --- | --- |
|     | $ZrO_2$ | ZrC | $ZrO_2$ | ZrC | Cu |
| OGB | 0.39% | 2.28% | 64 nm | 80 nm | 75 nm |
| AIF | 0.56% | 2.21% | 96 nm | 93 nm | 71 nm |

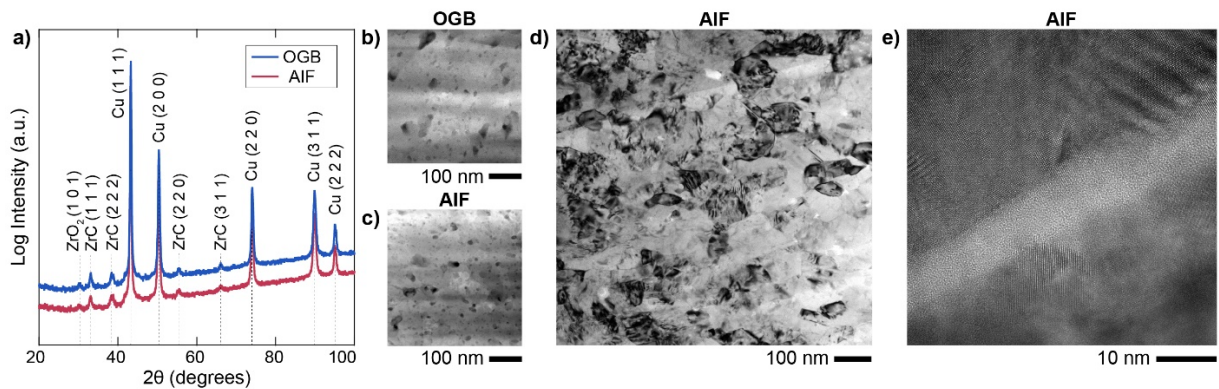

**Figure 1.** (a) XRD profiles from OGB and AIF samples of nanocrystalline Cu-Zr. The primary phase present is a Cu-rich face centered cubic structure, while small amounts of oxide and carbide impurity phases ($ZrO_2$ and $ZrC$) are also present. STEM-HAADF images show a similar size and distribution of impurity phases in both (b) OGB and (c) AIF samples. (d) Typical grain structure of the Cu-Zr powders, as viewed in BF-STEM imaging mode. (e) High resolution TEM image of an AIF in the quenched Cu-Zr sample.

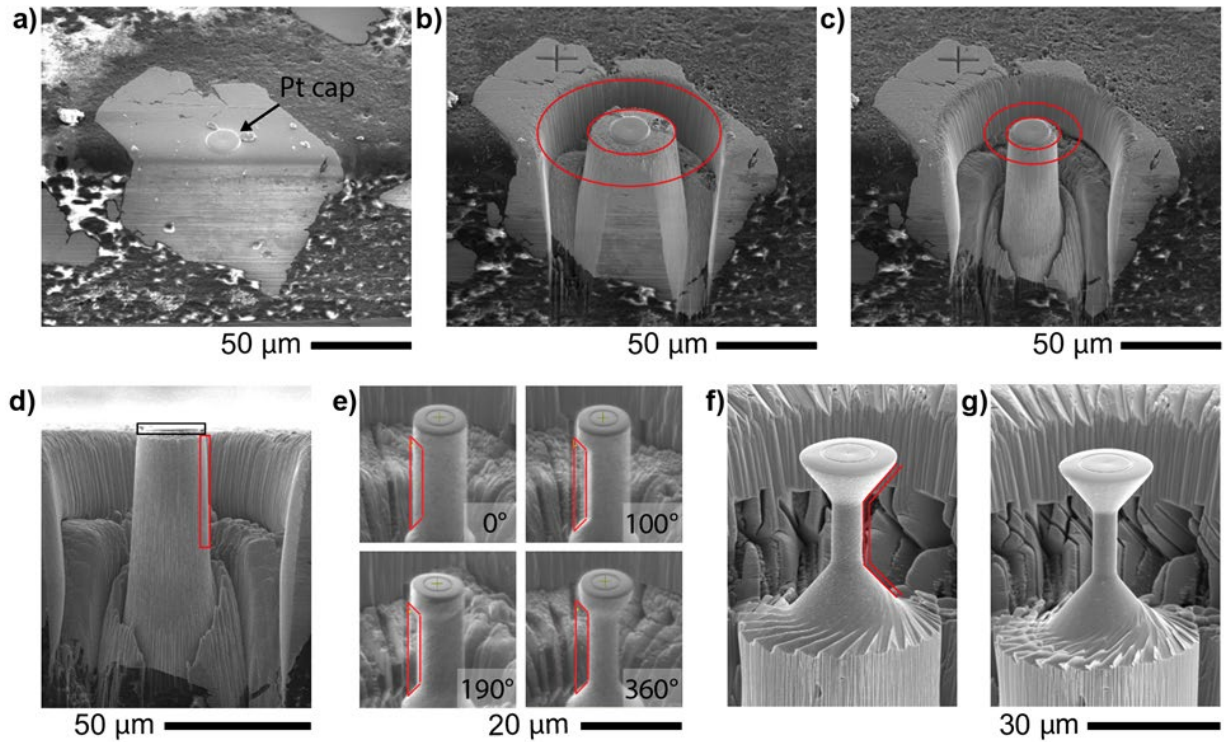

**Figure 2.** (a) A protective Pt cap (black arrow) is deposited on a Cu-Zr powder particle embedded in a graphite/epoxy mixture with the top and side surfaces exposed. (b) A high current (65 nA) annular milling step creates the rough pillar shape, followed by (c) a lower current (7-15 nA) annular milling step which increases the pillar aspect ratio and refines the pillar shape. (d) The pillar is viewed from the side and the Pt cap is leveled (black box) to prepare the Pt surface to create a fiducial mark for pattern matching. Pillar taper and asymmetry is removed during the first pass of lathe-milling (red box), (e) The pillar is shaped into a tensile specimen during the second pass of lathe milling, as demonstrated through the ion-beam view of the pillar at multiple rotation angles. (f) The tensile shape is refined during the penultimate pass and (g) the final pass provides a final surface polish.

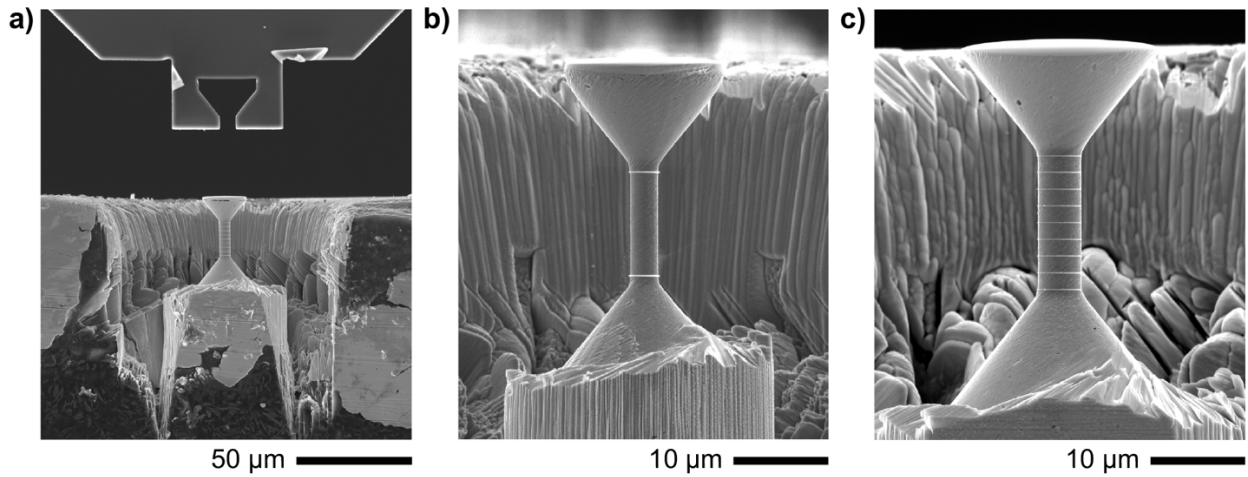

**Figure 3. (**a) A Cu-Zr powder particle embedded in the stiff epoxy matrix that has been shaped into a micron-scaled tensile sample, while a tensile grip is fashioned from the nanomechanical testing system force sensor head using FIB. (b) Two electron-beam deposited Pt markers are used to track the global strain on the smaller diameter sample set. (c) Nine Pt markers are used to track both global and local strains on the larger diameter sample set.

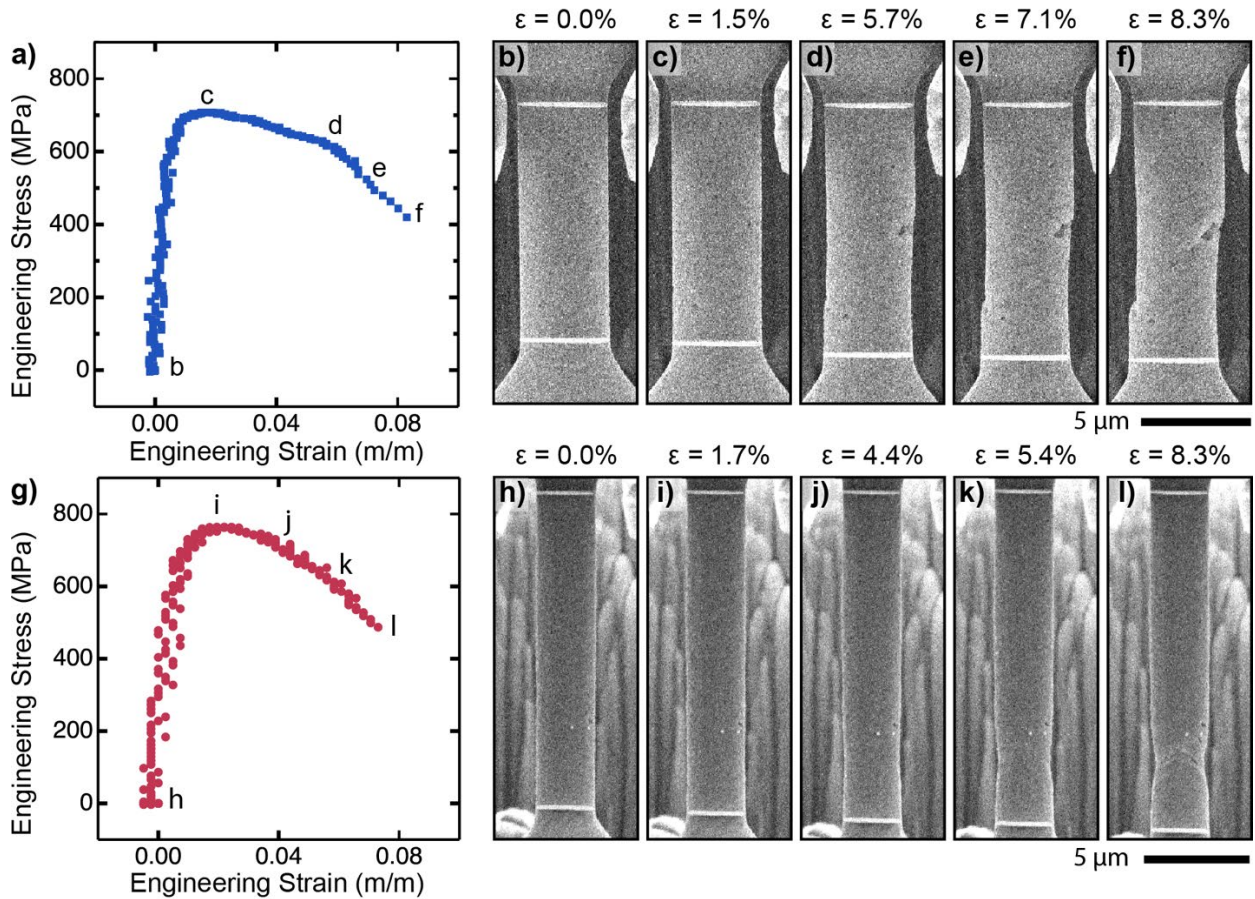

**Figure 4.** Stress-strain curves of (a) an OGB sample and (g) an AIF sample, with letter labels indicating the corresponding SEM image to the right. Parts (b) and (h) show the undeformed geometries, (c) and (i) show the sample state at maximum uniform elongation (strain at ultimate tensile stress), and (d)-(f) and (j)-(l) show the progression of failure in each sample either by shear-dominated failure for the OGB sample or by diffuse necking for the AIF sample.

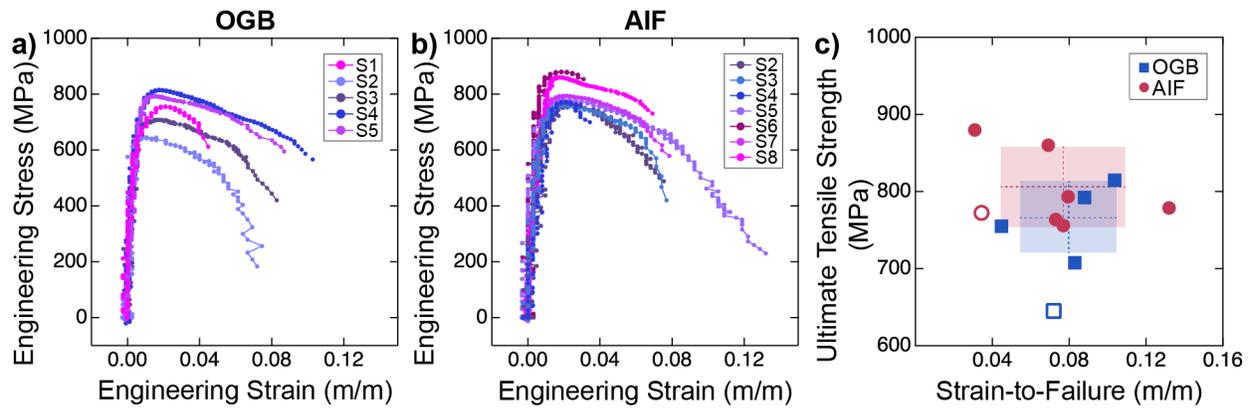

**Figure 5.** Stress-strain curves from (a) five OGB samples and (b) seven AIF samples. (c) Compiled ultimate tensile strength and strain-to-failure values for all samples tested in this study. Horizontal and vertical lines indicate the average ultimate tensile strength and strain-to-failure values, respectively, for each sample type. The open data points from each set were not included in these calculations as they were too small to rule out external size effects (see detailed discussion in text). Shaded boxes outline the regions associated with one standard deviation from the average values. Average ultimate tensile strength was $767 \pm 47$ MPa for the OGB samples and $805 \pm 52$ MPa for the AIF samples, while average strain-to-failure was $8.0 \pm 2.5\%$ for the OGB samples and $7.7 \pm 3.2$ % for the AIF samples.

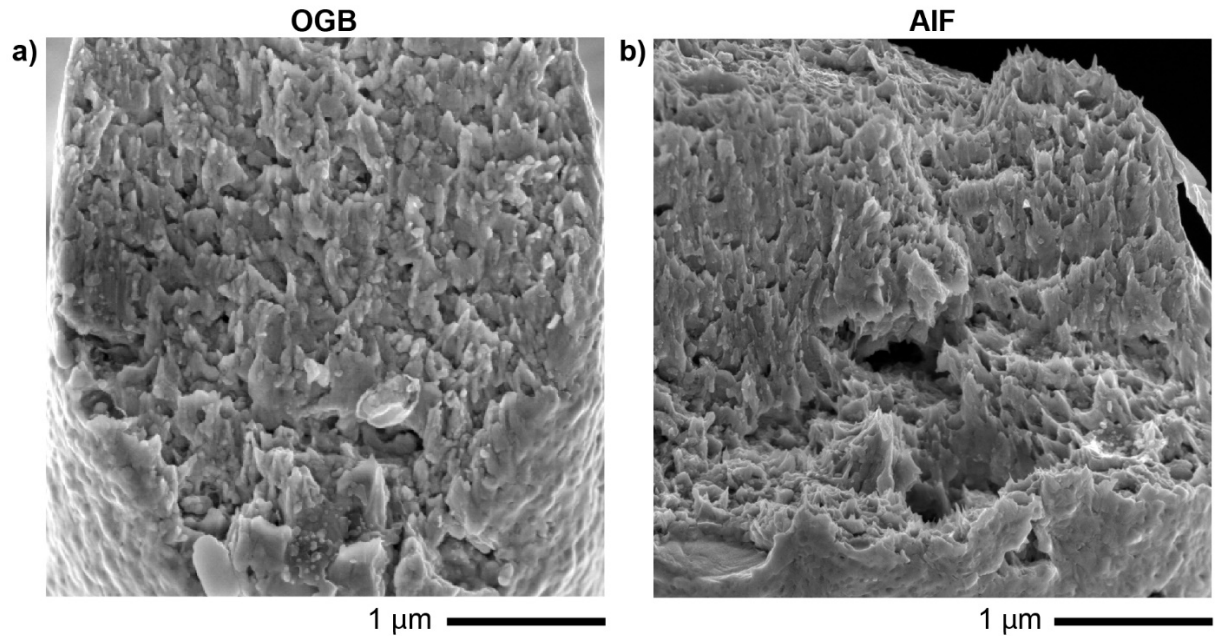

**Figure 6.** Fracture surfaces of (a) an OGB sample and (b) an AIF sample taken at 5 kV at an angle of 40°, with both images presented at the same magnification. The fracture surfaces of the AIF samples demonstrated more elongated cup-and-cone features, which are most visible in the bottom right of (b).

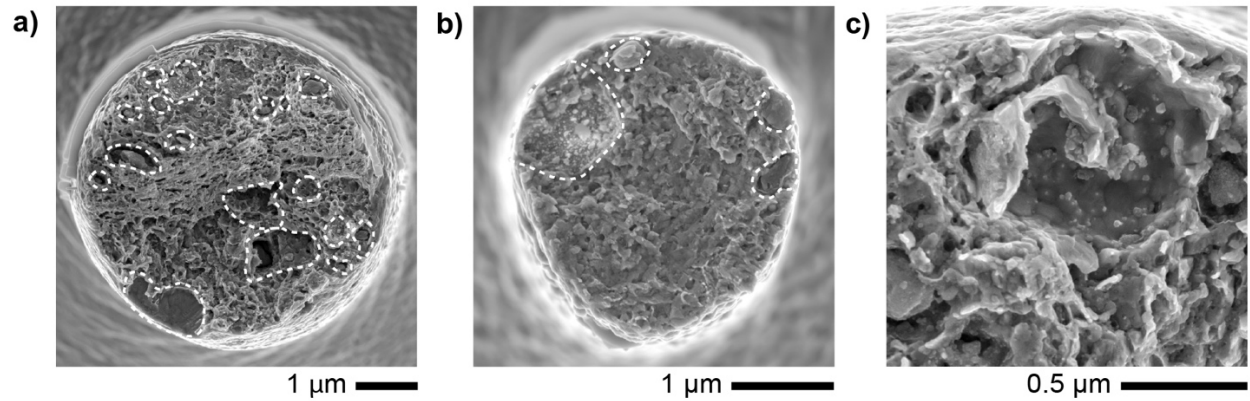

**Figure 7.** Fracture surfaces from two AIF-containing samples, to demonstrate the effect of processing defects such as pores or impurity phases. In both cases, processing defects outlined with dashed white lines act as fracture initiation sites. In (a), the individual defects are small compared to the sample dimensions, so the applied load was able to be supported by other material and a reasonably large strain-to-failure was achieved. In (b), a pore that was large relative to the sample size led to early failure. (c) Magnified image of a defective region showing the range of defect types and sizes found in fracture surfaces, from nanometer-sized particles to micron-sized porous cavities.

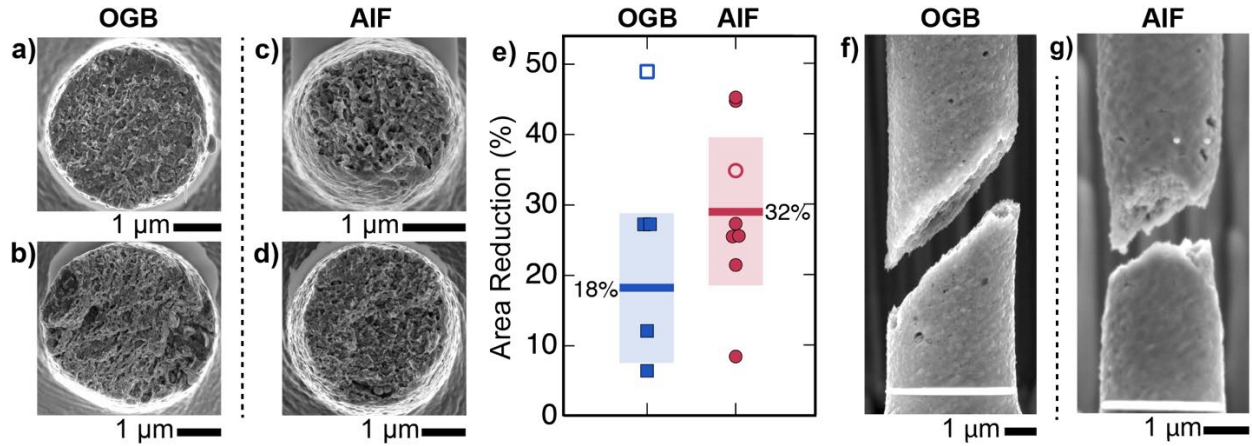

**Figure 8.** OGB sample fracture surfaces with measured area reductions of (a) 12% and (b) 6%, as well as AIF sample fracture surfaces with much larger measured area reductions of (c) 45% and (d) 26%. (e) Summary of the cross sectional area reduction for all samples, with average values indicated by solid horizontal lines and standard deviations shown by shaded areas. Side views of samples post-failure for the (f) OGB and (g) AIF materials.

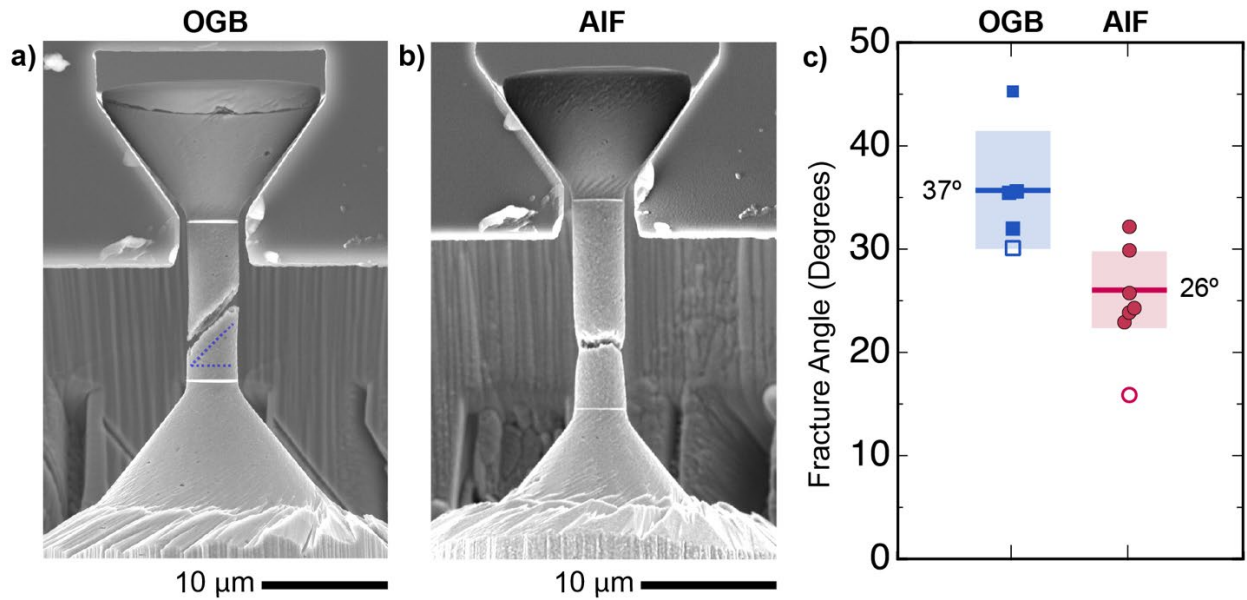

**Figure 9.** (a) An OGB sample and (b) an AIF sample after failure displaying shear-dominated and necking failure, respectively. (c) Summary of fracture angle for each sample type, as determined by fitting a plane to the fracture surfaces from a side view. The average failure angles (solid horizontal lines) show that OGB samples fail at an angle that is steeper than the AIF samples, providing another point of reference between the sample sets.

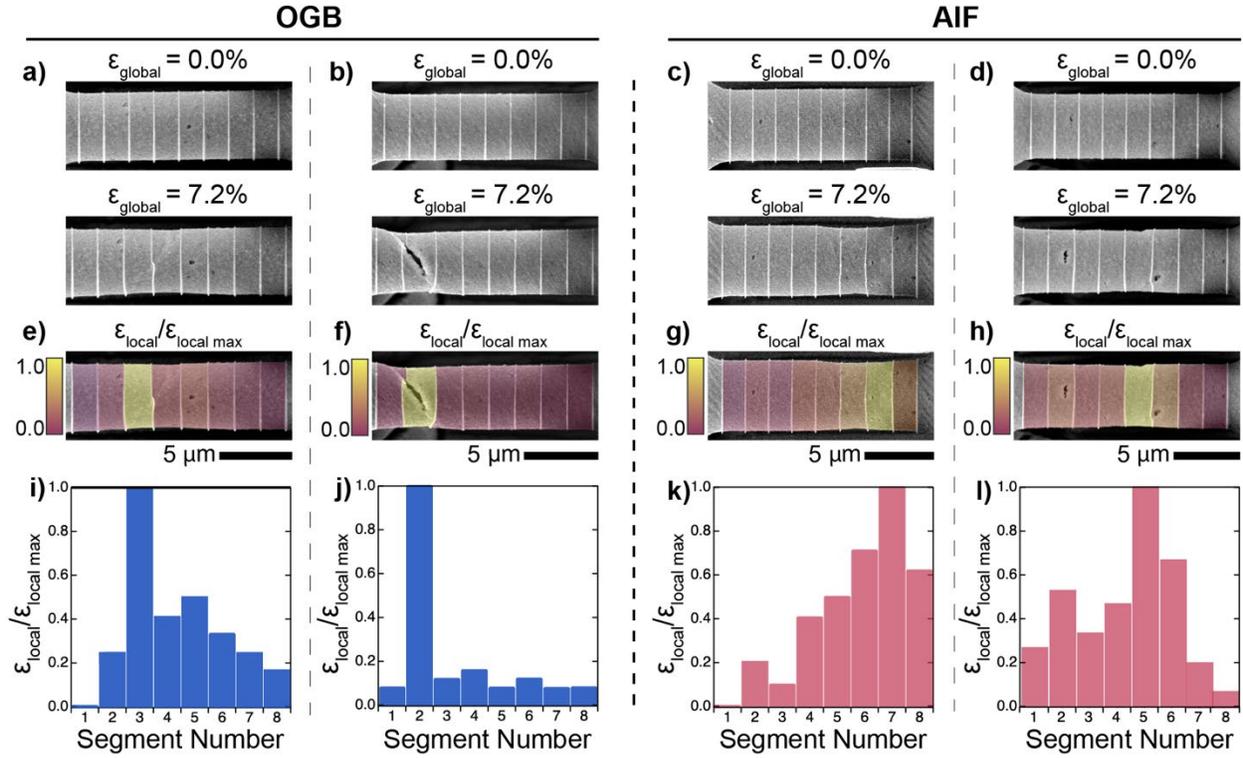

**Figure 10.** (a)-(d) In situ SEM images of undeformed (Row 1) and deformed (Row 2) samples. (e)-(h) Local strain distribution across the gauge section of micro-tensile samples at a globally strained state of 7.2%. $\varepsilon_{local}$ is the local strain in each segment, and $\varepsilon_{local\ max}$ is the maximum local strain in across the entire sample at 7.2% global strain, meaning their ratio provides a normalized metric to compare across samples. (i)-(l) The spatial distribution of normalized local strain in each sample. Segment 1 is the left most segment (at the base of the tensile test piece) and Segment 8 indicates the right most segment (top of tensile test piece). The OGB samples demonstrate strong strain localized to one segment along the gauge section, whereas AIF samples have a more homogeneous and well-distributed strain profile.

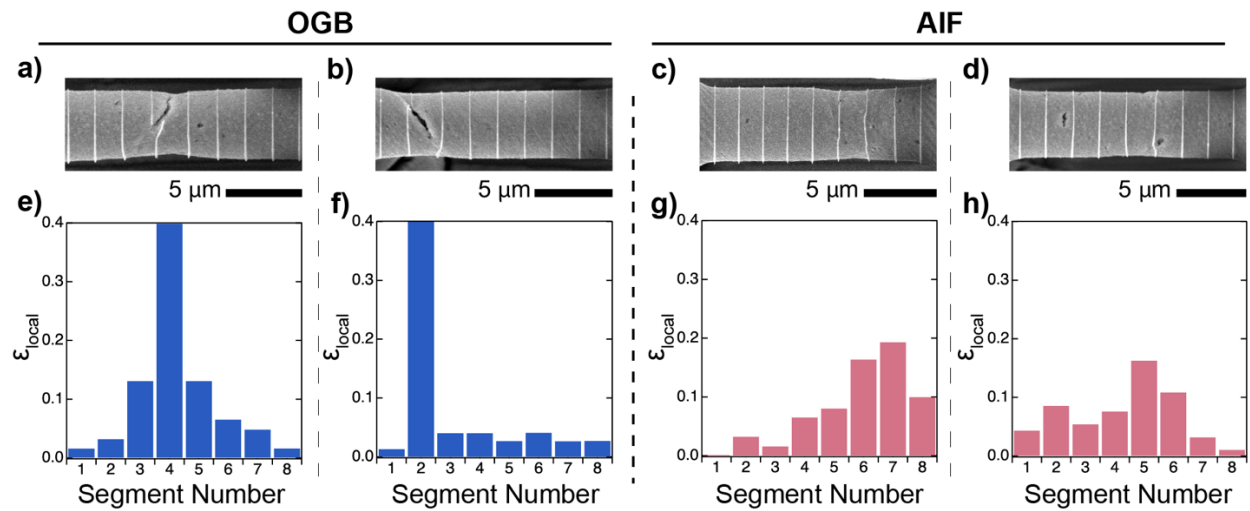

**Figure 11.** (a)-(d) In situ SEM images of deformed samples immediately prior to failure. Visible cracking is found in OGB samples while diffuse necking is seen in the AIF samples. (e)-(h) Local strain along the tensile axis, demonstrating that plasticity is more homogeneously distributed in the AIF sample.